# Engineered Defects to Modulate Fracture Strength of Single Layer MoS$_2$: An Atomistic Study

Rafsan A.S.I. Subad,[a] Tanmay Sarkar Akash,[a] Pritom Bose,[a] and Md Mahbubul Islam*[b]

[a]Department of Mechanical Engineering, Bangladesh University of Engineering and Technology, Dhaka-1000, Bangladesh

[b]Department of Mechanical Engineering, Wayne State University, 5050 Anthony Wayne Drive, Detroit, MI- 48202, USA

*Corresponding Author. Tel.: 313-577-3885; E-mail address: mahbub.islam@wayne.edu

**Abstract**

We use classical molecular dynamics (MD) simulations to investigate the mechanical properties of pre-cracked, nano-porous single layer MoS$_2$ (SLMoS$_2$) and the effect of interactions between cracks and pores. We found that the failure of pre-cracked and nano-porous SLMoS$_2$ is dominated by brittle type fracture. Bonds in armchair direction show a stronger resistance to crack propagation compared to the zigzag direction. We compared the brittle failure of Griffith prediction with the MD fracture strength and toughness and found substantial differences that limit the applicability of Griffith's criterion for SLMoS$_2$ in case of nano-cracks and pores. Next, we demonstrate that the mechanical properties of pre-cracked SLMoS$_2$ can be enhanced via symmetrically placed pores and auxiliary cracks around a central crack and position of such arrangements can be optimized for maximum enhancement of strengths. Such a study would help towards strain engineering based advanced designing of SLMoS$_2$ and other similar transition metal dichalcogenides.

# 1. INTRODUCTION

$MoS_2$ is one of the most widely studied layered transition metal dichalcogenides (TMDs). Single layer $MoS_2$ ($SLMoS_2$) is an excellent super-conductor with a direct bandgap of 1.8 eV,[1] this specific property of $SLMoS_2$ is very promising to overcome the limitations of gapless graphene as such it can potentially play a tremendous role on widespread application areas in the field of energy conversion[2] and storage[3], photodetectors, integrated logic circuits, transistor[4], sensors[5], hydrogen evolution reactor (HER)[6], and optoelectronics[7]. In addition to excellent electronic properties, $SLMoS_2$ also possesses some impressive mechanical properties including a well-balanced stiffness and flexibility that make it a favourable candidate for filler in nano-materials like nano-porous filter for water desalination[8].

The possibility of frequent exposure to elevated temperature and harsh chemical environment during manufacturing and real-life applications makes $SLMoS_2$ susceptible to the growth and evolution of defects. Multi-atom defects such as cracks and pores can cause mechanical properties to compromise severely. Such defects are inevitable in many situations and often they are deliberately impressed upon the nanostructures to harness desired properties, especially in terms of electrical and optical properties and even to enhance mechanical stability for some nanomaterials. For example, $MoS_2$ monolayers originated from chemical vapour deposition (CVD) usually exhibit much lower carrier mobility compared to the mechanically exfoliated $SLMoS_2$. It is because, generally, various defects in 2-D sheets[9–11] are prevalent in chemical growth processes. In the applications of TMDs as nano-catalysts and dry lubricants, defects due to radiation damage are unavoidable and can cause severe structural damages if not carefully fabricated and maintained. $SLMoS_2$ filters with nano-pores have emerged very recently exploiting their promising performance for membrane separation and pores which play the central role in this regard[12]. Suppression of defect formation is not possible or desired in such cases. One way to leverage the effort would be to engineer the defects and adopting a stress-engineering approach to minimize such defect induced damages and enhance the mechanical properties, if possible. Such enhancements were previously documented for several nanomaterials and it may be a curious research question if such an approach is suitable for TMD family of nanomaterials, including SLMoS$_2$. Therefore, it is essential to investigate the effect of size and behaviour of deliberate poring and cracking on the mechanical properties of SLMoS$_2$ not only to predict and prevent the mechanical failure but also to investigate if such defects can be exploited to enhance the mechanical properties itself.

Due to the widespread availability in nature, bulk MoS$_2$ is a well-studied layered TMD. However, for the case of single to few-layer MoS$_2$, the amount of study is still wanting. Bertolazzi et al.[13] exfoliated single and double layer MoS$_2$ and measured the in-plane elastic modulus and the failure strength. Their measurement showed the strength of SLMoS$_2$ to be close to the theoretical strength limit of Mo-S covalent bonds. Li[14] performed ab initio calculations to measure the strength of SLMoS$_2$ and reported that the failure mechanism is attributed to the out of plane relaxation of atoms, different from non-buckled honeycomb structure of graphene. Moreover, distinct defect formation, reconstruction and their stability have been observed in case of SLMoS$_2$[9]. Bao et al.[15] used classical molecular dynamics (MD) simulations to investigate the crack propagation mechanism and fracture toughness of SLMoS$_2$ and found that energy release rate in the fracture process decreases with increasing initial crack length, crack angle and temperature. They reported crack tip blunting due to stress concentration immediately before fracture propagation. Sulfur vacancy defects in MoS$_2$ was reported to alter the crack propagation path of SLMoS$_2$, resulting in an enhanced fracture toughness. Wang et al.[16] introduced patterned array of precisely controlled sub-nanometer (nm) pores down to 0.6 nm on SLMoS$_2$ and showed the stability of pore adjacency closer than 5 nm. Although few studies were conducted concerning the mechanical properties and fracture mechanism aspects of pristine and cracked SLMoS$_2$, there is a knowledge gap for porous SLMoS$_2$, their elastic properties and fracture mechanism under tensile loading. Additionally, there is not enough computational or experimental study that helps to answer if mechanical properties of pre-cracked SLMoS$_2$ sheets can be enhanced by prefabricated array of cracks or pores, as can be speculated from similar healing phenomenon previously found in Graphene[17] and Silicene[18].

In this work, we studied the mechanical properties of centrally-cracked and centrally-pored SLMoS$_2$ at temperatures of 1K and 300K varying the crack size from 1.5nm to 5.5nm and diameter of the pore from 1nm to 3nm at a fixed strain rate. We compared the fracture strength and stress intensity factor against the Griffith theory. Additionally, geometries with central-crack and central-pore and the temperature effect on the crack-pore and primary crack- auxiliary crack combinations have also been investigated by varying temperature from 1K to 500K. To find out the interactions between crack-pore and primary crack-auxiliary cracks, we determined stress-strain relationship, stress and displacement distributions for different auxiliary crack and pore positions around the central primary crack, for both armchair and zigzag directional loading at 1K and 300K temperature. Finally, we also elucidate the fracture mechanism in all the cases studied here.

## 2. METHODOLOGY

We created 30 nm x 30 nm SLMoS$_2$ sheets using a MATLAB[19] script. The effective thickness of the nanosheet is taken to be 0.61 nm[20]. It has been reported that in order to avoid size effects of finite dimension, the minimum dimension (either length or width) of nanosheet must be at least ten times the maximum half crack length present in the sheet.[21,22] Therefore, to study the effect of crack length on the fracture strength of SLMoS$_2$ nanosheet, five cracks of different lengths (approximately 1.5 nm, 2.5 nm, 3.5 nm, 4.5 nm and 5.5 nm) were introduced in the center of the nanosheet both in armchair and zigzag direction and sheet dimensions were chosen according to the ten-fold rule. Additionally, to investigate the effect of pore size on the fracture strength of SLMoS$_2$ nanosheet, five pores of different diameters (approximately 1 nm, 1.5 nm, 2 nm, 2.5 nm and 3 nm) are introduced at the centre of the structure. Pores are created by deleting atoms from the pristine structure.

Furthermore, to examine the crack-pore (CP) interactions and primary crack-auxiliary crack (PC-AC) interactions, one primary crack is kept fixed at the centre and four pores or four auxiliary cracks are symmetrically introduced at different positions on each side of the central crack. A schematic illustration of the primary, auxiliary crack/pore positions are presented in Supporting Information (SI) Figure S1. The **sizes** of both the main crack and auxiliary cracks are 3.5 nm and the diameter of the pore is 1 nm. Because the pores and auxiliary cracks are symmetric in positions about the origin, the position of pore is designated by a single coordinate in the positive quadrant. We note that in our simulations armchair and zigzag directions are aligned to the x- and y-axis, respectively. For auxiliary cracks we studied the effect of their placements by stretching both along the cracks and away from it in the transverse directions. However, pores were found to extend their stress relaxation zone to a limited space. Therefore, it is not essential to probe their effect as they are placed transversely away from the crack plane. Herein, for pores, we studied the effect of pore positions along the crack only. Pore and auxiliary crack positions are given in Table S1 (see supplementary).

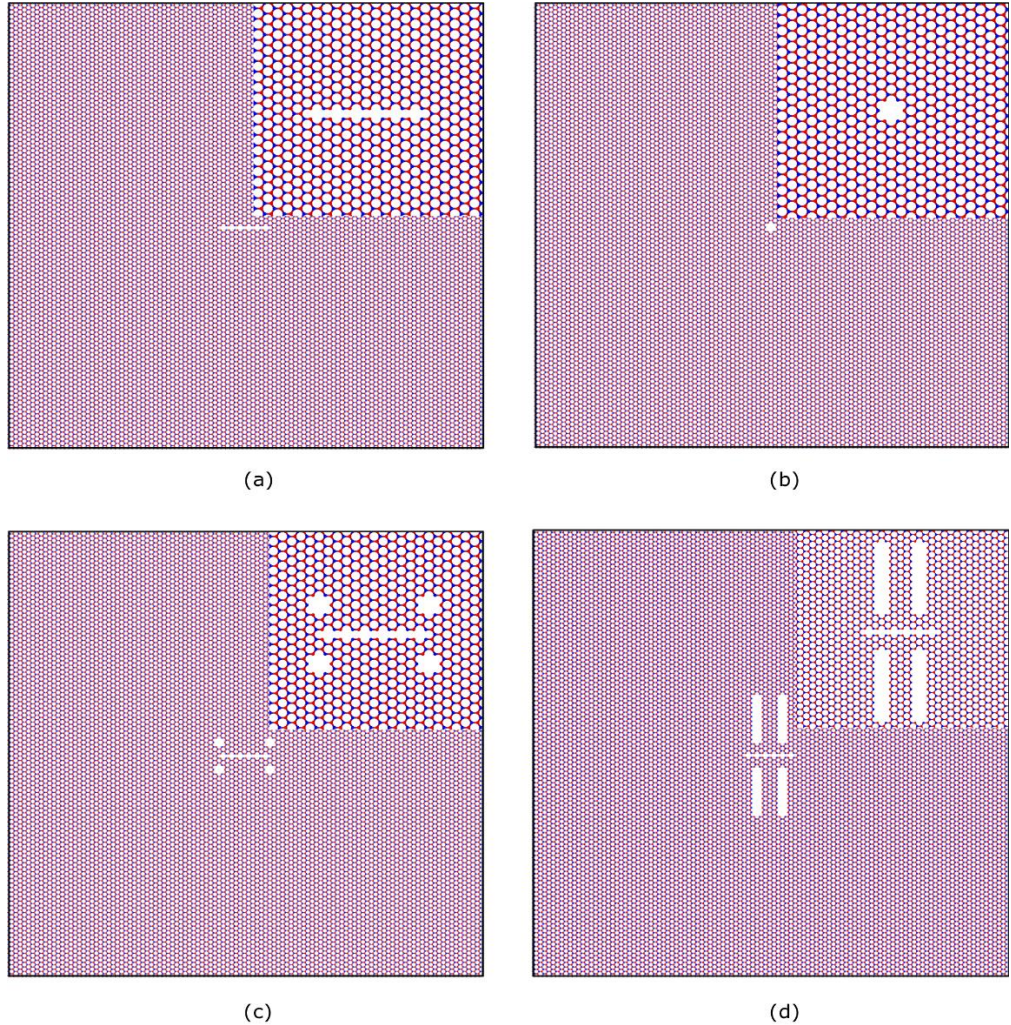

Figure 1- Atomic structure of SLMoS$_2$. (a) pre-crack in armchair direction (b) central nano-pore (c) armchair crack with auxiliary pores (d) armchair crack with auxiliary cracks. Atom color: S, red; Mo, blue.

We used MD simulations to determine mechanical properties and to study the fracture mechanism of SLMoS$_2$. LAMMPS [23] software package is used to perform all the simulations. Periodic boundary condition is maintained in the planar directions (x and y), while z-direction is non-periodic. We used a time step of 1 fs. The structure was minimized using a conjugate gradient algorithm (CG). Next, the structure was relaxed using an NVE ensemble followed by an NPT ensemble at atmospheric pressure and the target temperature. Pressure and temperature damping constants are 1, and .1 ps respectively in both XX and YY directions. We used a constant tensile strain rate of $10^9$ s$^{-1}$. Although the applied strain rate is relatively higher compared to the practical cases, it is widely used in MD simulations to study mechanical properties of nanoscale materials.[24,25] OVITO[26] software package is used for visualization. Atomic stress-strain behaviour is obtained by deforming the simulation box uniaxially and calculating the average stress over the structure. Atomic stress was calculated on the basis of the definition of virial stress. Virial stress components are calculated using the following relation[27]:

$$\sigma_{virial} = \frac{1}{\Omega} \sum_i \left( -m_i \dot{u}_i \otimes \dot{u}_i + \frac{1}{2} \sum_{j \neq i} r_{ij} \otimes f_{ij} \right) \quad (1)$$

where the summation is over all the atoms occupying the total volume $\Omega$, $\otimes$ indicates the cross product, $m_i$ is the mass of atom i, $r_{ij}$ is the position vector of the atom, $\dot{u}_i$ is the time derivative which indicates the displacement of an atom with respect to a reference position, and $f_{ij}$ is the interatomic force applied on atom i by atom j. Engineering strain is implied by the term strain and it can be calculated as,

$$\varepsilon = \frac{l - l_0}{l_0} \quad (2)$$

Here, $l_0$ is the undeformed length of the box where $l$ is the instantaneous length.

In this study, we used Stillinger–Weber (SW) potential by *Jiang et al.*[20] to define the interatomic interactions. The SW potential consists of a two body term and a three body term describing the bond stretching and bond breaking, respectively. The mathematical expressions are as follows

$$\Phi = \sum_{i<j} V_2 + \sum_{i>j<k} V_3 \quad (3)$$

$$V_2 = A e^{\left[\frac{\rho}{r - r_{max}}\right]} \left( \frac{B}{r^4} - 1 \right), \quad (4)$$

$$V_3 = K\varepsilon e^{\frac{\rho_1}{r_{ij} - r_{max\,ij}} \frac{\rho_2}{r_{ik} - r_{max\,ik}}} (\cos\theta - \cos\theta_0)^2 \quad (5)$$

Here $V_2$ and $V_3$ the two body bond stretching and angle bending terms accordingly. The terms $r_{max}$, $r_{max\ ij}$, $r_{max\ ik}$ are cut-offs and, $\theta_0$ is the angle between two bonds at equilibrium configuration. $A$ and $K$ are energy related parameters that are based on Valence Force Field (VFF) model. $B$, $\rho$, $\rho_1$, and $\rho_2$ are other parameters that are fitted coefficients. These parameters and their corresponding value can be found in ref. 14.

## 3. METHOD VALIDATION

In order to validate the approach used in the study, we applied uniaxial tension to both armchair and zigzag directions of the pristine 30 nm x 30 nm single layer $MoS_2$ at 1K and 300K temperatures. The corresponding stress-strain curves are shown in Figure 2. The calculated fracture strength and Young's modulus are then compared with literature.[20] The comparison between the results is shown in Table 2. It is evident from the comparison that the results predicted in the present study are in good agreement with the literature.[20,28]

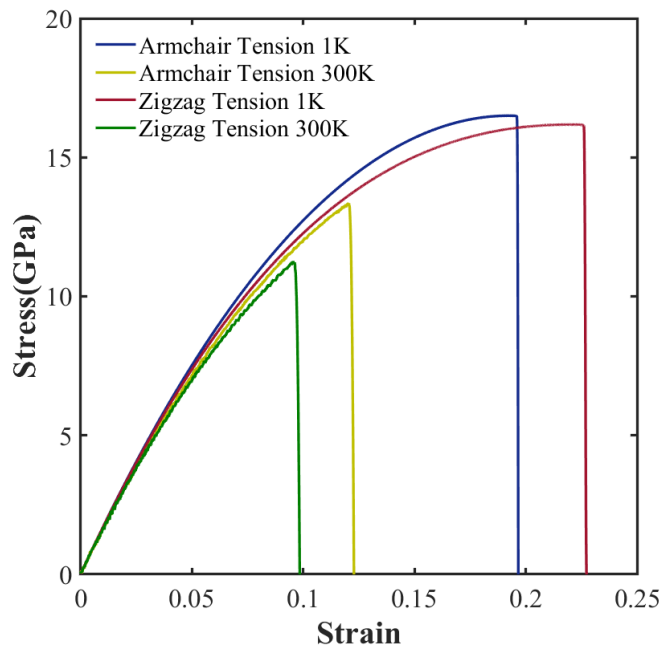

Figure 2- Stress-strain relationship for pristine $SLMoS_2$

**Table 1-** Validation of our calculation with the existing literature

| Mechanical Properties | This Study | | Previous Study[20] | | Previous Study[28] |
|---|---|---|---|---|---|
| | armchair loading | zigzag loading | armchair loading | zigzag loading | |
| Young's Modulus (GPa) | ~158.1 | ~159.7 | 165.5 | 167.0 | 174±43 120±30 (N/m) |
| Ultimate Tensile Stress (GPa) | 16.5 (1K) 13.3 (300K) | 16.2 (1K) 11.3 (300K) | ~16.8 (1K) ~13.1 (300K) | ~15.9 (1K) ~10.8 (300K) | |
| Fracture Strain (%) | 19.6 (1K) 12.3 (300K) | 22.7 (1K) 9.8 (300K) | ~19.9 (1K) ~12.1 (300K) | ~21.8 (1K) ~9.4 (300K) | |

We also investigate the possibility of phase transition during the failure of SLMoS2 under tensile loading and we found no transition from 2H to 1T' phases during the deformation. The SLMoS2 has the highest energy gap between 2H and 1T' phases among other TMDs, which alludes that the phase transition through uniaxial straining is less probable.[29–31] Our observation is consistent with previous DFT [32] and nano-indentation study using atomic force microscopy.[13] However, conflicting reports such as two stage deformation and phase transitions are also available in literature. [33,34] We note that, previous result[32] confirms that, for MoS$_2$, 1T phase is a metastable state which may be destroyed easily by vacancy or long wavelength perturbations. Metastable states often indicate an intermediary for structural transition, however, it is not still clear if before dissipation through cracks, MoS$_2$ can actually go through similar transition as observed for graphene to form triangularly bonded atomic chain after crack propagation at a critical strain larger than ultimate strain.[35] We further compare the fracture-to-failure energy between MD simulations (without structural transition) and DFT (with structural transition) by calculating the critical energy release rate, G$_c$. The Figure S11 shows variation of $G_c$ with different initial crack lengths. Our calculated G$_c$ of 3.1eV/Å slightly overestimates the DFT result of 2.21eV/Å,[36] and REBO MD result of (2.5eV/Å),[37] indicating a lesser brittle characteristic compared to other results. Overall, our calculated results compare well with the previously reported data.

## 4. RESULTS AND DISCUSSION

**4.1 Stress-strain relationships for nano-porous SLMoS$_2$**

The fracture strength of porous SLMoS$_2$ largely depends on the pore size. Such dependency can be found in Figure 3 for five nanopores with diameters between 1-3 nm for both armchair and zigzag loading directions. Detailed stress-strain relations are presented in supplementary Figure S2 (b),(d) (in armchair loading) and (a),(c) (in zigzag loading), where the calculated stress is the nominal stress in the SLMoS$_2$. Here, single nanopore is located at the centre of the sheet. Thus, the size of the pore is varied to find out the size effect of nanopore on the SLMoS$_2$ sheet. The tensile strength decreases as the size of the nanopore increases which follows a similar trend as nanoporous graphene sheets. [38] Increasing void space enhances the probability of fracture and causes the earlier failure of the material. The tensile strength and fracture tensile strain in zigzag direction are larger than the armchair as we have discussed later and such result is consistent with previous studies.[39]

Figure 3 (a) and (b) also show the comparison between the computed fracture stress from MD simulations and Griffith theory of brittle fracture, respectively in zigzag, and armchair loading conditions. From Griffith theory of brittle failure,[40] fracture stress [41] can be computed from the equation :

$$\sigma_f = \sqrt{\frac{2\gamma E}{\pi a_0}} \qquad (6)$$

Where γ is the surface energy for 3-D materials and edge energy for 2-D materials, E is the Young's modulus, $a_o$ is the initial radius of the pore.

The surface energy is computed from the bond energy at 0 K. Bond energy is potential energy per bond. From the bond energy, surface energy is computed using the following equation:

$$\gamma = \frac{E_b}{2t\Delta L} \quad (7)$$

Where is $E_b$ is the bond energy, t is the sheet thickness and $\Delta L$ is the crack length extension due to bond breaking. $\Delta L$ is taken as $\sqrt{3}r_p$ (armchair loading condition) and $1.5r_p$ (zigzag loading condition) [42], where

$$r_p = \sqrt{r^2 - h^2} \quad (8)$$

(r and h denote the bond length and buckling height at equilibrium at zero strain at 0 K).

Equation (6) can be re-written as follows

$$\sigma_f \sqrt{\pi a_0} = \sqrt{2\gamma E} \quad (9)$$

The quantity $\sigma_f \sqrt{\pi a_0}$ is termed as fracture toughness. According to Griffith, the right-hand side of the equation is only material dependent. As it is constant for a certain material, the left-hand side of the equation should also be constant. Supplementary Figure S3 shows the variation of fracture toughness with crack length. Griffith theory is based on brittle fracture within continuum and elastic framework, however, MoS$_2$ has discrete atomic structure. Also Griffith theory was developed for bulk material and it is nearly impossible to maintain Griffith crack pattern in MoS$_2$ sheet. These contribute to variation in edge energy that eventually leads to the variation of fracture toughness.

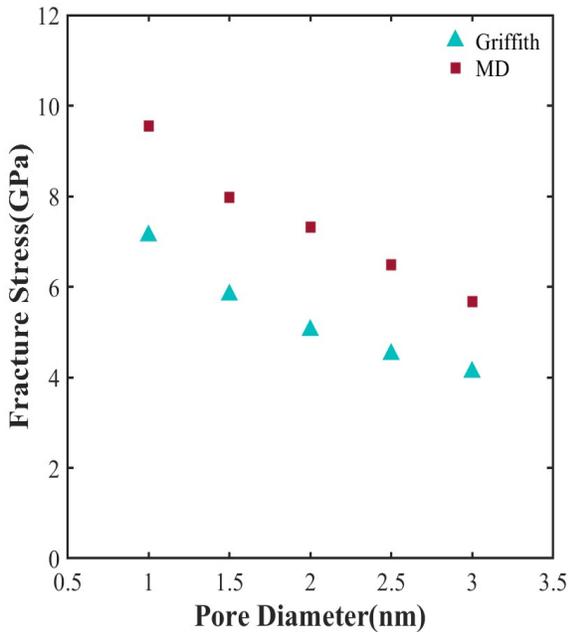
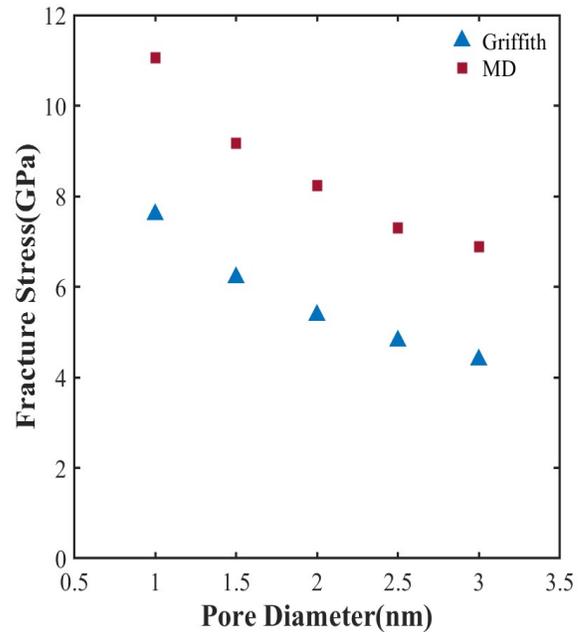

(a)            (b)

Figure 3- Change of the fracture stress with the change of pore diameter estimated by MD and Griffith formula under uniaxial tension along the (a) armchair (b) zigzag direction.

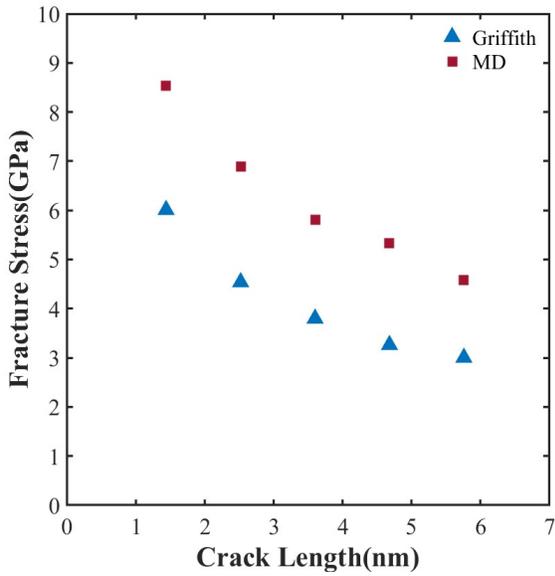
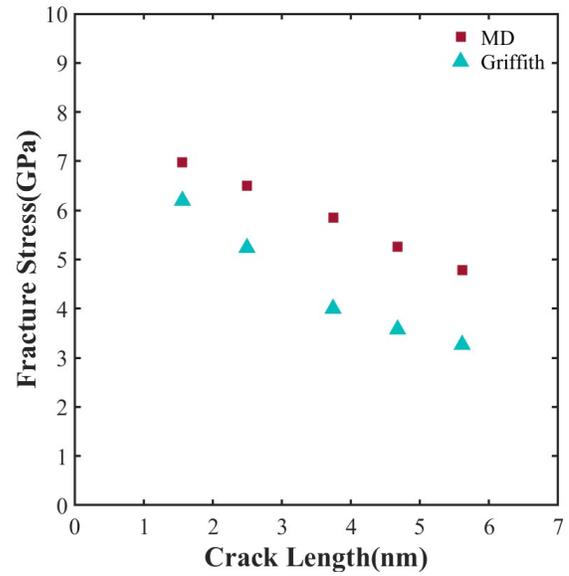

(a)            (b)

Figure 4- Evolution of the fracture stress with the change of crack length estimated by MD and Griffith theory under uniaxial tension along the (a) armchair (b) zigzag directions.

**4.2 Stress-strain relationships for pre-cracked SLMoS$_2$**

We studied the effect of central crack length on the ultimate stress of SLMoS$_2$ nanosheet at 1K and 300K. The stress-strain relationship of MoS$_2$ nanosheet for zigzag and armchair loading is shown in supplementary Figure S4 (a), (c) and (b), (d), respectively. Previous studies reported the failure of typical 2-D material is brittle in nature.[43,44] It is evident that the ultimate stress and failure strain decrease with the increase in crack length for both zigzag and armchair loading directions. This trend is similar for both low (1K) and high (300K) temperature. This is because, with the increase in crack length, stress localization around the crack tip also increases[40,45] as such decreases the materials ultimate stress and failure strain. In the elastic regime, both the pre-cracked and their corresponding pristine structure show almost identical uniaxial stress-strain curves in both zigzag and armchair loading conditions. Therefore, the Young's modulus of pre-cracked nanosheet is approximately equal to that of the pristine one. However, for same crack length, ultimate stress and failure strain are higher for loading in zigzag than the armchair direction. This is because in case of zigzag loading condition, the bonds are either at 30° or 60° angle with loading direction but in case of armchair loading, the bonds are parallel to the loading direction. Therefore, the bonds can stretch more in zigzag loading than the armchair loading direction before they reach critical bond length. This results in higher ultimate stress and failure strain in zigzag loading condition. Comparisons of fracture stress and fracture toughness derived from our simulations with those obtained from calculations based on Griffith's criterion are presented in Figure 4 and Figure S5. It is evident from the figures that, Griffith's theory underestimates the MD results for nanoporous $MoS_2$.

**4.3 Crack Enhancement with prefabricated pores**

Depending on the position of pores around the central crack, fracture of the structure can be amplified or shielded. Figure 5 shows the stress-strain curve of the $SLMoS_2$ sheet with pores around the central crack for both zigzag and armchair loading conditions at 1K and 300K temperatures. Figure 6 demonstrates the variation of ultimate stress for different positions of pores around the central crack. It is evident from Figure 6 that for loading in both the directions, the effect of pore on fracture for Pore Position 1 (see Figure S1) is almost negligible. The strength of the material can be improved for the Pore Position 2 to 5. To further understand the reason behind the increment of the ultimate stress of the sheet, we compare stress distribution of the pre-cracked sheet without pore and with pore at different positions both for zigzag (at 4.5% strain) and armchair (at 4.0% strain) loading conditions as shown in Figure 7(a) and 7(b), respectively. Figure 7(a)(i) and 7(b)(i), indicate that stress concentration near the crack tip is the primary reason of failure for pre-cracked $SLMoS_2$. Additionally, Figure 7(a)(ii) and 7(b)(ii) corroborate the claim that pore position 1 has a negligible effect on the fracture as it does not play a role to alter the stress distribution close to the crack tip region. Therefore, the crack tip remains stressed as the case without pores, and the fracture initiates before the pores are even stressed. Finally, 'Crack Shielding Effect' due to the presence of pores has been observed, and represented in Figure 7(a)(iii), and 7(b)(iv). The stress field near the crack tip with pores are distributed over a large area around the pores (which is shown in these two figures). Therefore, the average stress near the crack tip is reduced and thus the crack shielding effect[18] has been induced. Crack tips with pores near the crack edges were found to be less stretched for the same

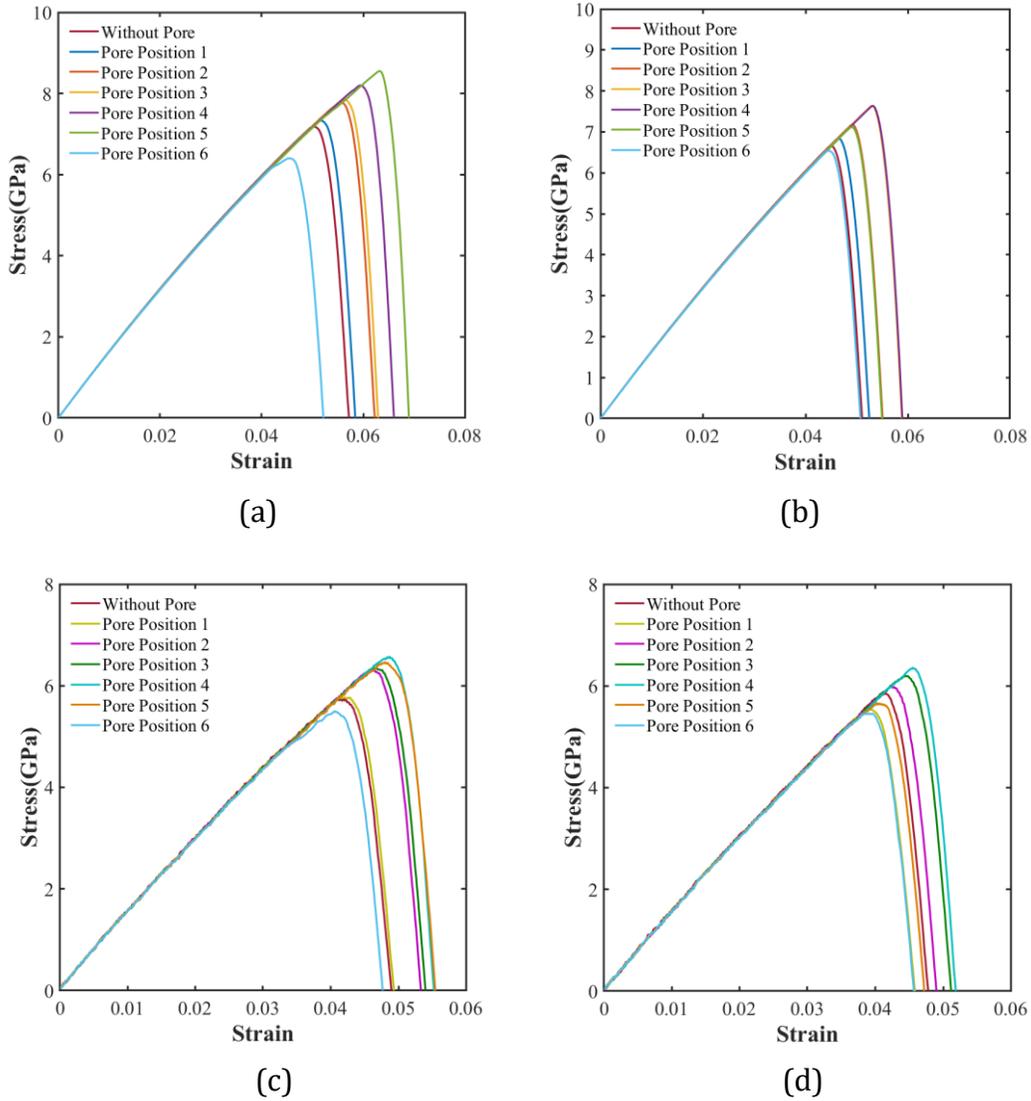

Figure 5- Stress-strain curves for crack-pore interaction in SLMoS$_2$ keeping a fixed crack length 3.5nm and pore diameter 1nm (a) crack in armchair direction at 1K (b) crack in zigzag direction at 1K (c) crack in armchair direction at 300K (d) crack in zigzag direction at 300K

applied far-field strain (Figure 8) than without nearby pores which resists the early failure of the material. We observed any pores beyond pore Position 5, weakens the strength of the material. For pore position 6 and loading along zigzag direction (Figure 7(a)(iv)), the stress is distributed around the pores but the position of pores lies on the direction of crack propagation path and very adjacent to the tip that crack and pore get merged and then crack propagates through the pore weakening the material severely. The loading along armchair direction (Figure 7(b)(iv)),

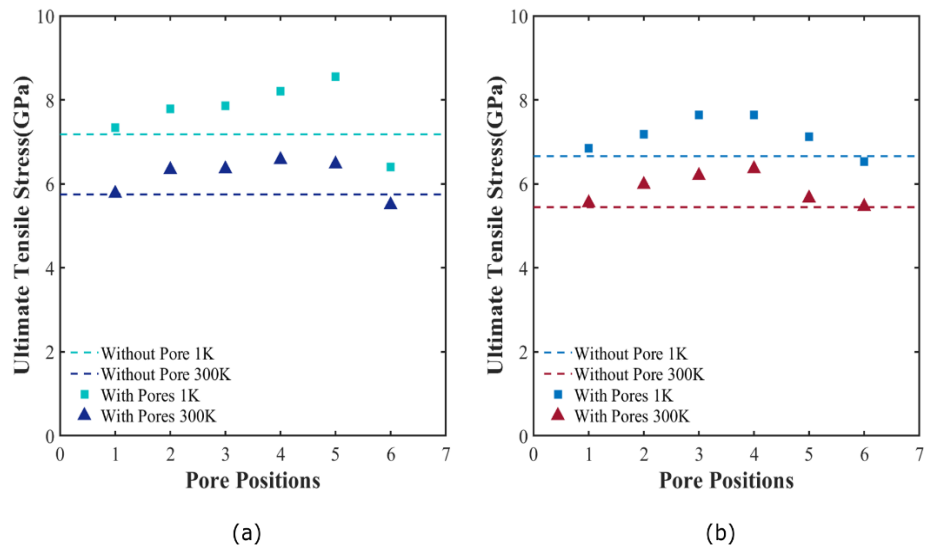

(a)                                        (b)

Figure 6- Crack enhancement with the help of pores (a) loading along zigzag direction (b) loading along armchair direction

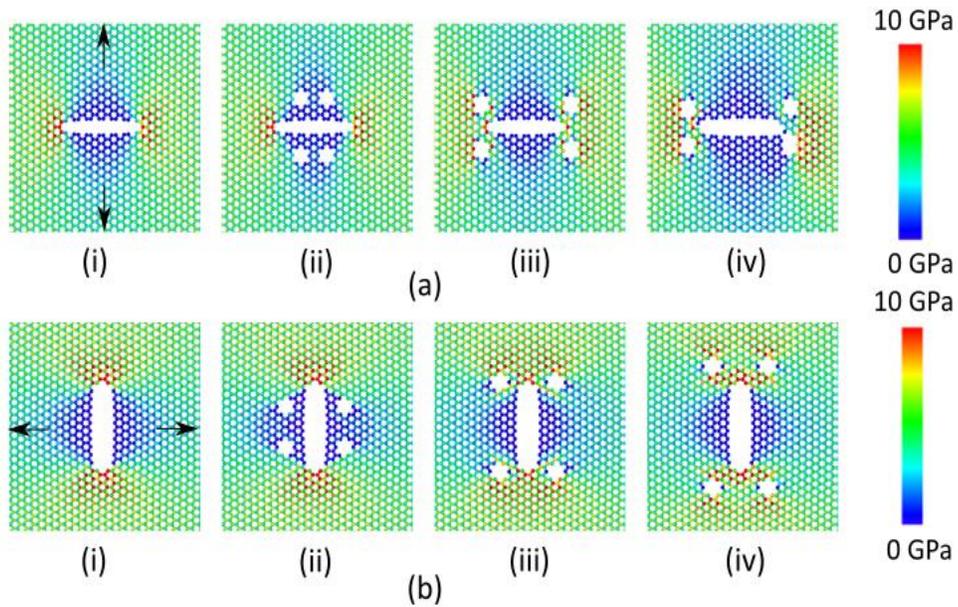

Figure 7- Stress distribution (a) for loading along zigzag direction (i) only crack (ii) pore position 1 (iii) pore position 5 (iv) pore position 6 at 4.5% strain and (b) for loading along armchair direction (I) only crack (ii) pore position 1 (iii) pore position 4 (iv) pore position 6 at 4% strain

the average stress near the crack tip increases due to the position of the stressed lower portion of the pores in the proximity of the crack tip and thus induces the stress amplification effect[46] as such early failure occurs for both the cases.

## 4.4 Crack Enhancement with Auxiliary Cracks

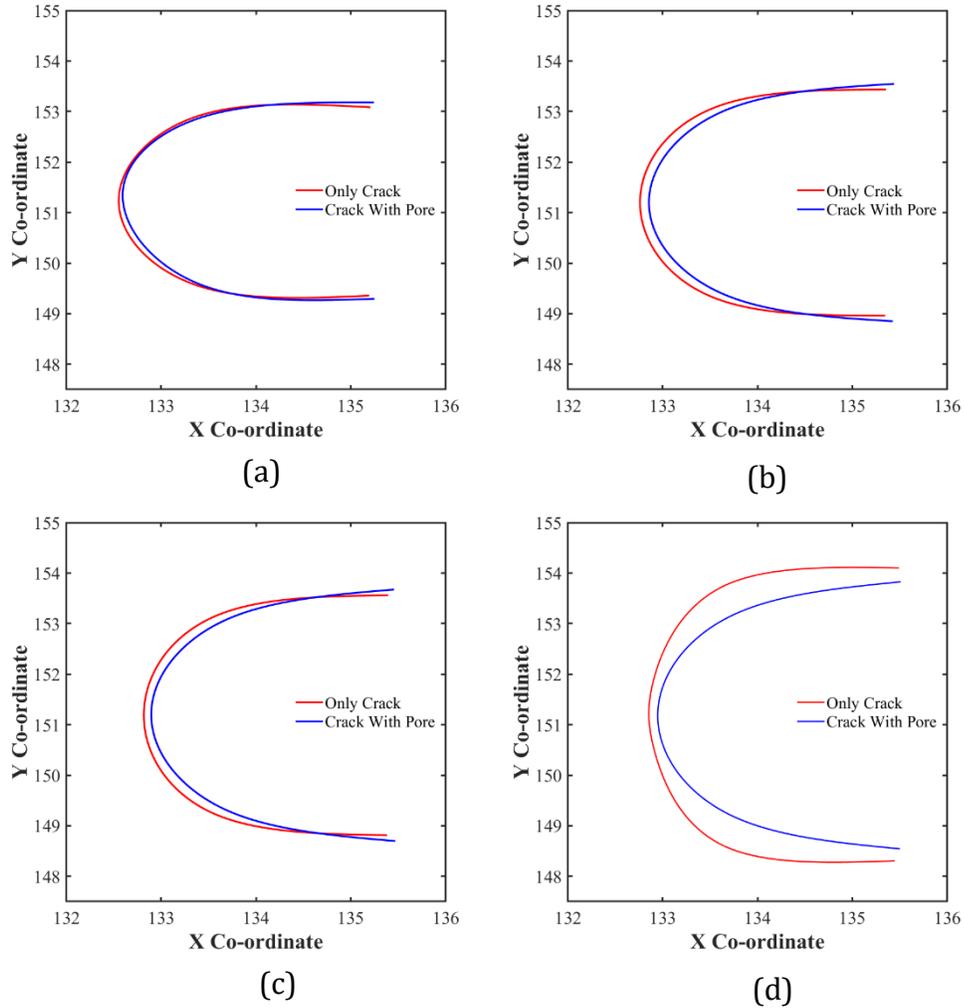

Figure 8- Crack tip positions at (a) 20% (b) 40% (c) 45% (d) 50% strain

To investigate the effect of auxiliary cracks on the propagation of main crack under tensile loading, auxiliary cracks are introduced around the central crack. The orientation of the auxiliary crack is kept perpendicular to the main crack or parallel to the loading condition. Crack length of the auxiliary cracks is kept same as the main crack (3.5 nm). For different samples with armchair and zigzag main cracks, the centre of the auxiliary cracks are, at first placed at different positions along the crack direction and at a fixed transverse position. For sample 1-6, auxiliary cracks are positioned increasingly towards the crack tips. For central armchair and zigzag crack

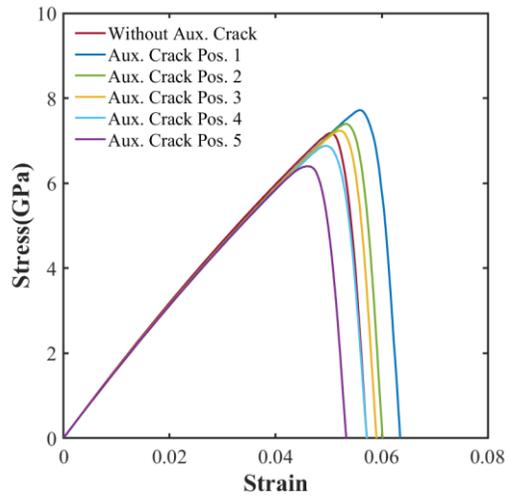
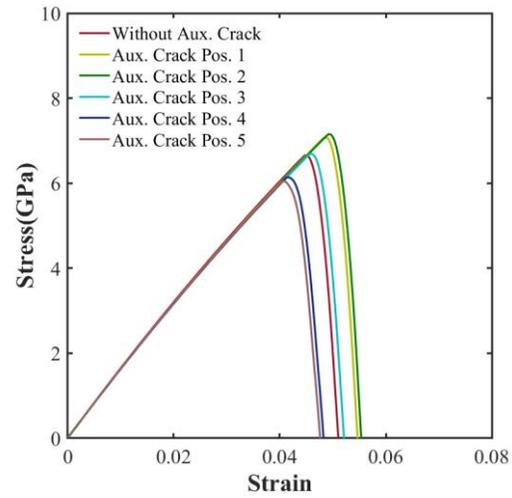

(a)  (b)

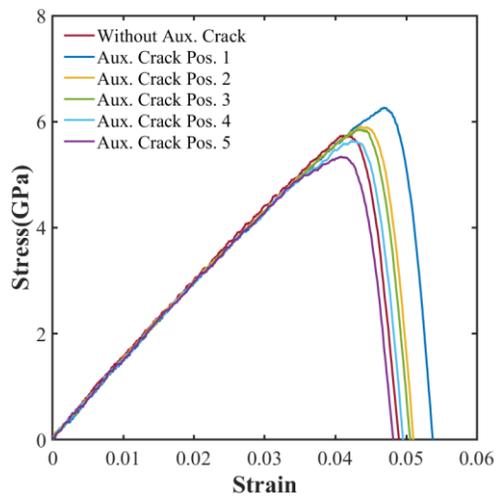
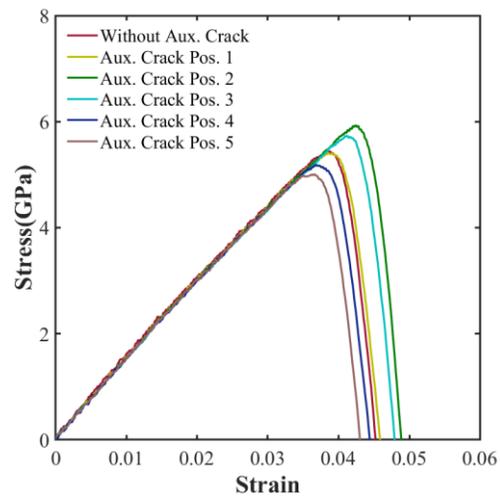

(c)  (d)

Figure 9- Stress-strain curve for enhancing crack with the help of auxiliary crack in SLMoS$_2$ keeping a fixed crack length 3.5nm for both (a) crack in armchair direction at 1K (b) crack in zigzag direction at 1K (c) crack in armchair direction at 300K (d) crack in zigzag direction at 300K

respectively, position 1 and 2 display an increased mechanical strength compared to all other auxiliary crack positions. For these two optimized positions along the crack, the auxiliary cracks are now placed increasingly away from the main crack along the transverse direction. Results (Figure 10) indicate that position 1 (for armchair main crack) and 2 (for zigzag main crack) remain to be superior with an enhanced mechanical strength considering all the positions of auxiliary crack.

For loading in armchair direction, the improvement of mechanical properties is found to be due to crack shielding mechanism. In the presence of auxiliary cracks parallel to the loading direction, unstressed region is created near the crack tip because of parallelism with loading

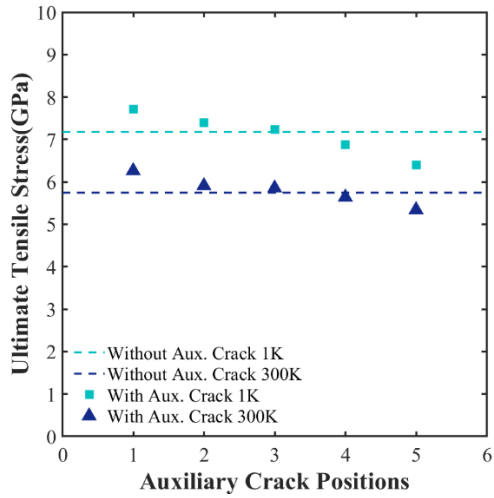
(a)

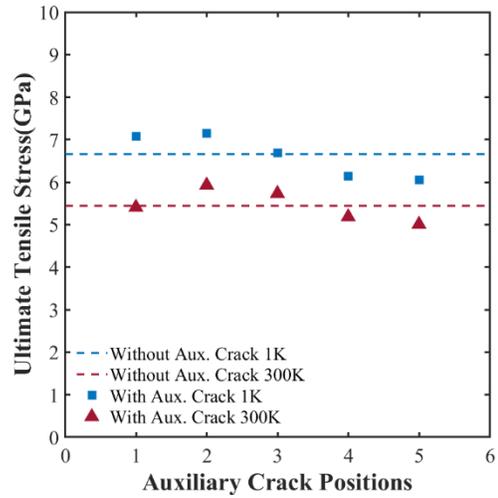
(b)

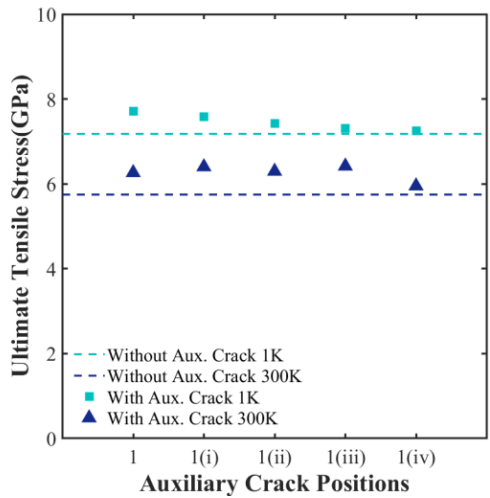
(c)

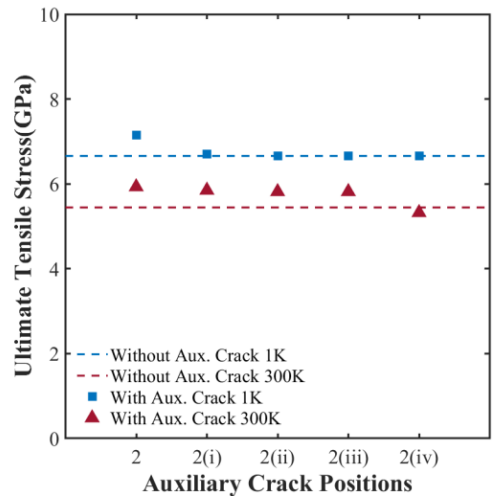
(d)

Figure 10- Crack enhancement by auxiliary cracks (a), (c) loading along zigzag direction and (b), (d) loading along armchair direction

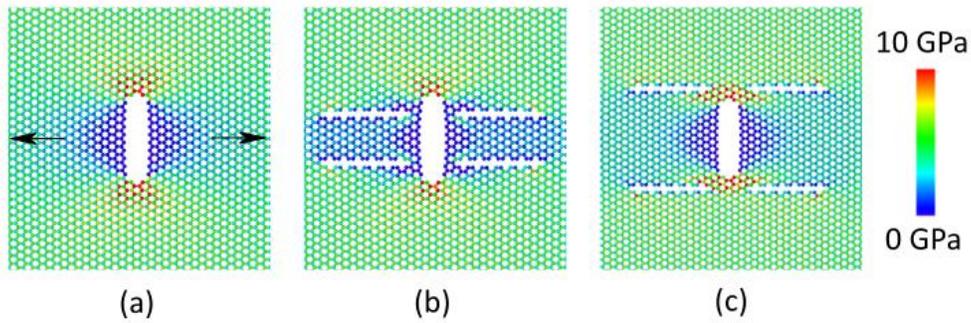

Figure 11- Stress distribution for loading along armchair direction (a) only crack (b) auxiliary crack position 1 (c) auxiliary crack position 5 at 3.8% strain. Arrow in figures shows the loading direction

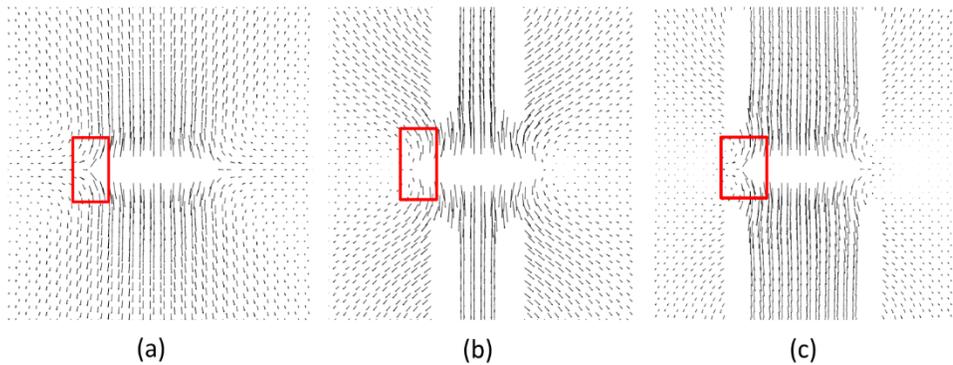

Figure 12- Displacement vectors for atoms around the crack tip before failure for SLMoS$_2$ containing (a) no auxiliary crack, (b) auxiliary crack at position 2 (c) auxiliary crack at position 5. Red colored box shows for (a) and (c), atomic movements at left crack tip are at an angle with the loading direction which hinders the bond rotation while for (b), atomic displacement are at larger angle than that of (a) and (c) (almost parallel) to the loading direction enabling the rotation of the bond at the crack tip.

direction. This unstressed region when superimposed with the stressed region near the main crack tip properly, reduces the stress near the main crack tip. As a result, when the auxiliary crack moves away from Position 2 in ± x direction (with respect to position 2), the unstressed region also moves along and therefore crack shielding effect gradually diminishes and at a point fracture amplification effect starts to take place (position 4,5). When the auxiliary crack is moved in y direction keeping the x coordinate fixed, the unstressed region is yet at the appropriate place but away from crack tip vertically. For the reason, weakening due to crack shielding effect is minimal.

For zigzag loading condition, crack shielding occurs due to bond rotation phenomenon at the crack tip. Figure 12 shows the bond rotation before failure near primary crack tip due to the presence of auxiliary cracks at different positions along the primary crack. Bond rotation was observed with atomic displacement analysis in OVITO. In the presence of auxiliary cracks, bonds near crack tip tend to rotate before stretching which delays the attainment of critical bond length. We speculate that the dangling atoms at the auxiliary crack tips favour such additional degrees of freedom. Therefore, the bonds require more energy to start stretching compared to the condition when stretching occurs in absence of auxiliary cracks. However, the atomic displacements at the tips are directed towards the loading direction enables this bond rotation[18] phenomenon, which happens in case of zigzag loading only.

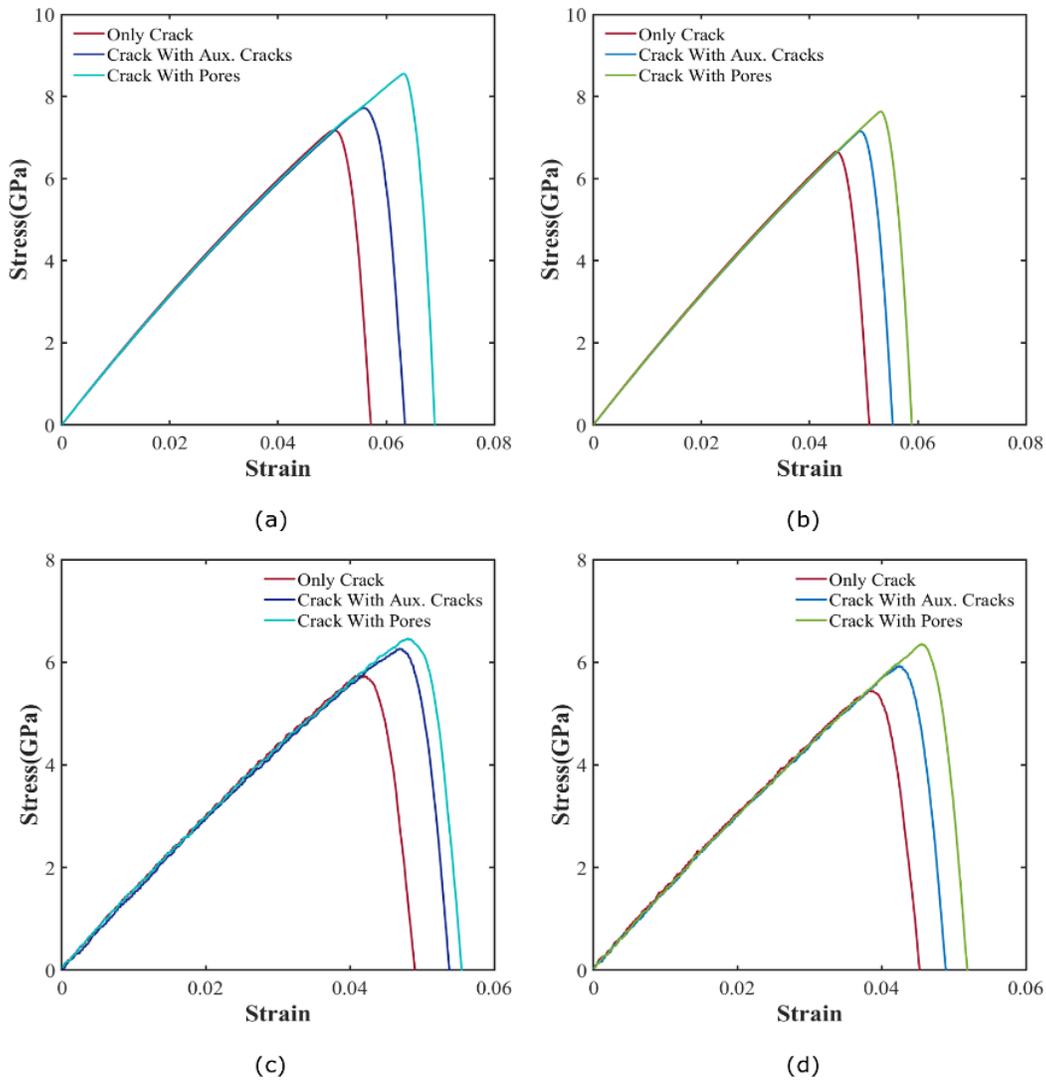

Figure 13- Stress-strain relationship for the comparison between crack-pore and crack-auxiliary crack system for (a) loading in zigzag direction at 1K (b) loading in armchair direction at 1K (c) loading in zigzag direction at 300K (d) loading in armchair direction at 300K

**4.5 Comparison between Crack Enhancement Techniques: Pores vs Auxiliary Cracks**

Figure 13 is a precise comparison of auxiliary crack and pore induced enhancement of ultimate strength at their optimized positions for the highest enhancement around the primary crack. It shows that the enhancement of strength is larger with pores around the central crack than with auxiliary cracks. On average, the ultimate tensile strength of the pre-cracked material has been enhanced almost 8% for main crack-auxiliary crack combination whereas 15% for the crack-pore combination. The increment is low compared to that of graphene in which a maximum 67% increase has been reported.[17] Graphene exhibits an extremely high fracture strength and higher stress concentration at the crack tip which widens the crack tip to reduce the stress concentration by placing cracks or pores around it. From this point of view, we can also predict that modulation the fracture strength of the TMDs will be less effective than graphene. The less stressed regions near the tip are spaced greater in case of auxiliary crack than pores. Therefore, the less stressed regions are close to the main crack with pores than auxiliary cracks. Furthermore, pores can alter the stress distribution more prominently than auxiliary cracks and, can share the stress near the main crack tip. Consequently, as pore has lower stress concentration, it contributes to higher ultimate stress.

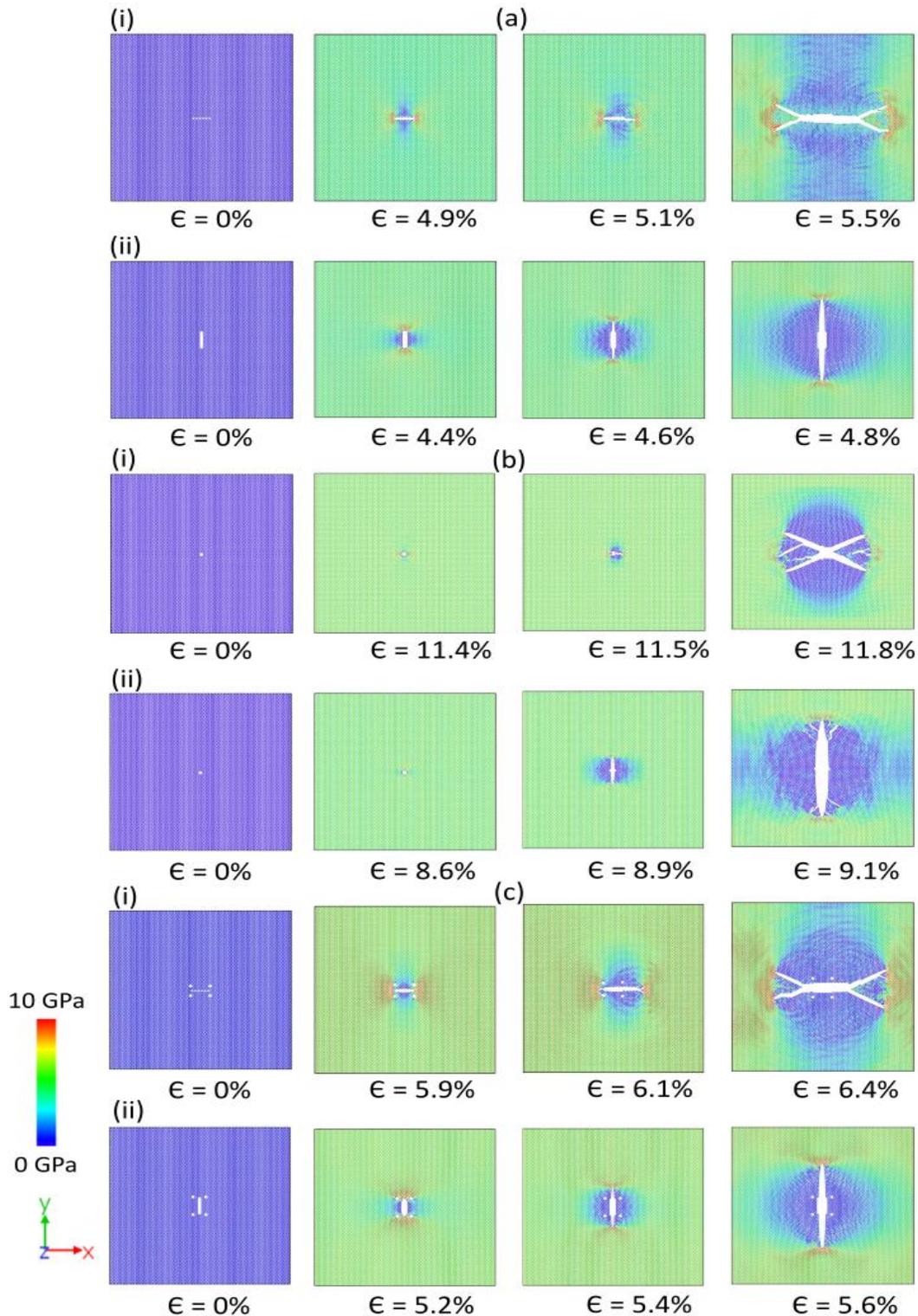

Figure 14- The stress distribution, deformation process and crack propagation of SLMoS$_2$ in (a) pre-crack (b) nano-pore (c) crack with auxiliary pores condition, tension along (i) zigzag (ii) armchair direction. The color bar shows the stress in GPa. Here, X and Y axis represents armchair and zigzag edges, respectively. The red dumbbell shaped zone near the crack tip is the plastic zone where permanent deformation takes place. As the strain increases, the area of this zone extends, facilitating fracture in the material.

## 4.6 Fracture Mechanism

We observed that the fracture mechanism for central pore, crack and crack with pore is almost analogous. For armchair crack, there are four (two at each side) symmetrically inclined bonds located at the crack tip that provide two potential crack propagation paths (±60° with x direction). Hence branching phenomenon is observed during crack propagation. However, in case of zigzag crack, there are two bonds at the crack tip that is perpendicular to the crack. Therefore, the crack propagation path is just an extension of the initial crack and no branching occurs.[47,48] This is one of the reasons why bonds in armchair direction resist fracture better than those in zigzag direction. Cracks with pores also follow the same mechanism except only when the crack and pore merge. At first few bonds break to facilitate crack tip like formation and then follow the mechanism of fracture of crack. Due to the bluntness of the crack during the failure of pore, the branches stretch along the zigzag direction. At the crack tip, the crack can propagate easily along any zigzag direction, but will have preference which is perpendicular to the load. As such there are few branches observed in case of failure of the pore along armchair loading.

## 5. CONCLUSION

In summary, we performed molecular dynamics simulations to investigate the mechanical properties and fracture behaviour of deliberately cracked and pored $SLMoS_2$ structure. We checked the result for different crack lengths and pore diameters. Increase in crack length and pore diameter leads to degradation of fracture strength, matching with the general trend of Griffith's brittle fracture model. However, Griffith's prediction underestimates the fracture stress, indicating the limitation of the model at the nanoscale. Temperature plays a significant role on the mechanical properties of cracked and pored samples. Failure stress and strain are both found to be deteriorated with the increasing temperature. We also studied the effect of crack and pore patterning around cracks to investigate their interaction with the central crack. Our study reveals that patterning of pores and auxiliary cracks can diminish or enhance the fracture toughness of cracked SLMoS$_2$ structure depending on their positions around the central crack. Pores and auxiliary cracks tend to relax the excessive stress concentration near the crack tips, delaying the fracture phenomenon. For zigzag loading with auxiliary cracks the crack shielding mechanism is primarily dictated by bond rotation near crack tip immediately before fracture. Finally, we made a comparison between crack enhancement techniques. It is found that pores can be used to enhance fracture strength more effectively than auxiliary cracks.

## 6. CONFLICTS OF INTEREST

There are no conflicts to declare.

## 7. ACKNOWLEDGEMENT

The authors of this paper would like to convey their thankfulness to Multiscale Mechanical Modeling and Research Network (MMMRN) group of BUET for their technical support. M.M.I acknowledges the support from Wayne State University startup funds.

# Supplementary Information for

# Engineered Defects to Modulate Fracture Strength of Single Layer MoS$_2$: An Atomistic Study

Rafsan A.S.I. Subad,[a] Tanmay Sarkar Akash,[a] Pritom Bose,[a] and Md Mahbubul Islam*[b]

[a]Department of Mechanical Engineering, Bangladesh University of Engineering and Technology, Dhaka-1000, Bangladesh.

[b]Department of Mechanical Engineering, Wayne State University, 5050 Anthony Wayne Drive, Detroit, MI- 48202, USA

*Corresponding Author. Tel.: 313-577-3885; E-mail address: mahbub.islam@wayne.edu

## SUPPLEMENTARY FIGURES

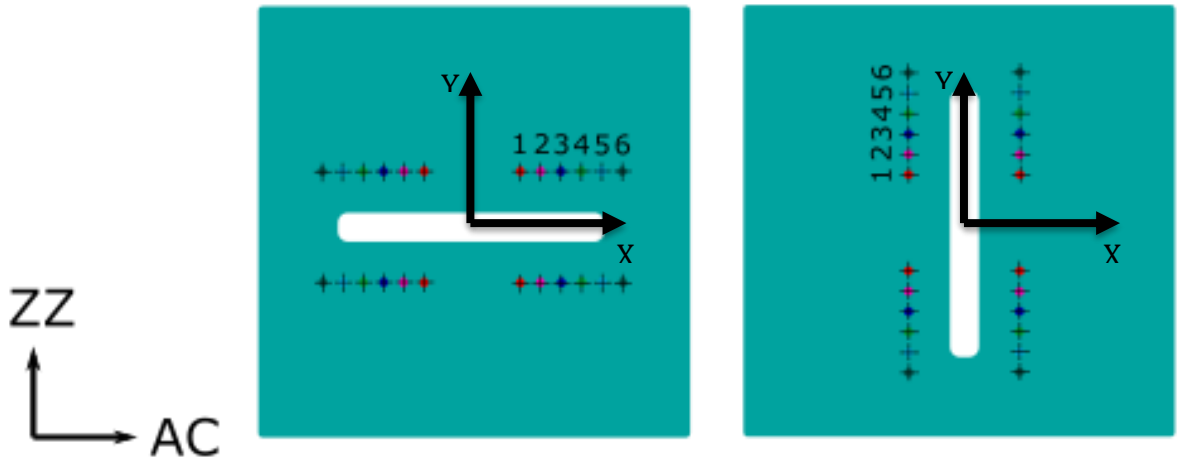

Figure S1- A schematic of the relative arrangements of the centers of auxiliary cracks/pores around the central crack. Six positions for symmetric auxiliary crack/pores are indicated in a single quadrant

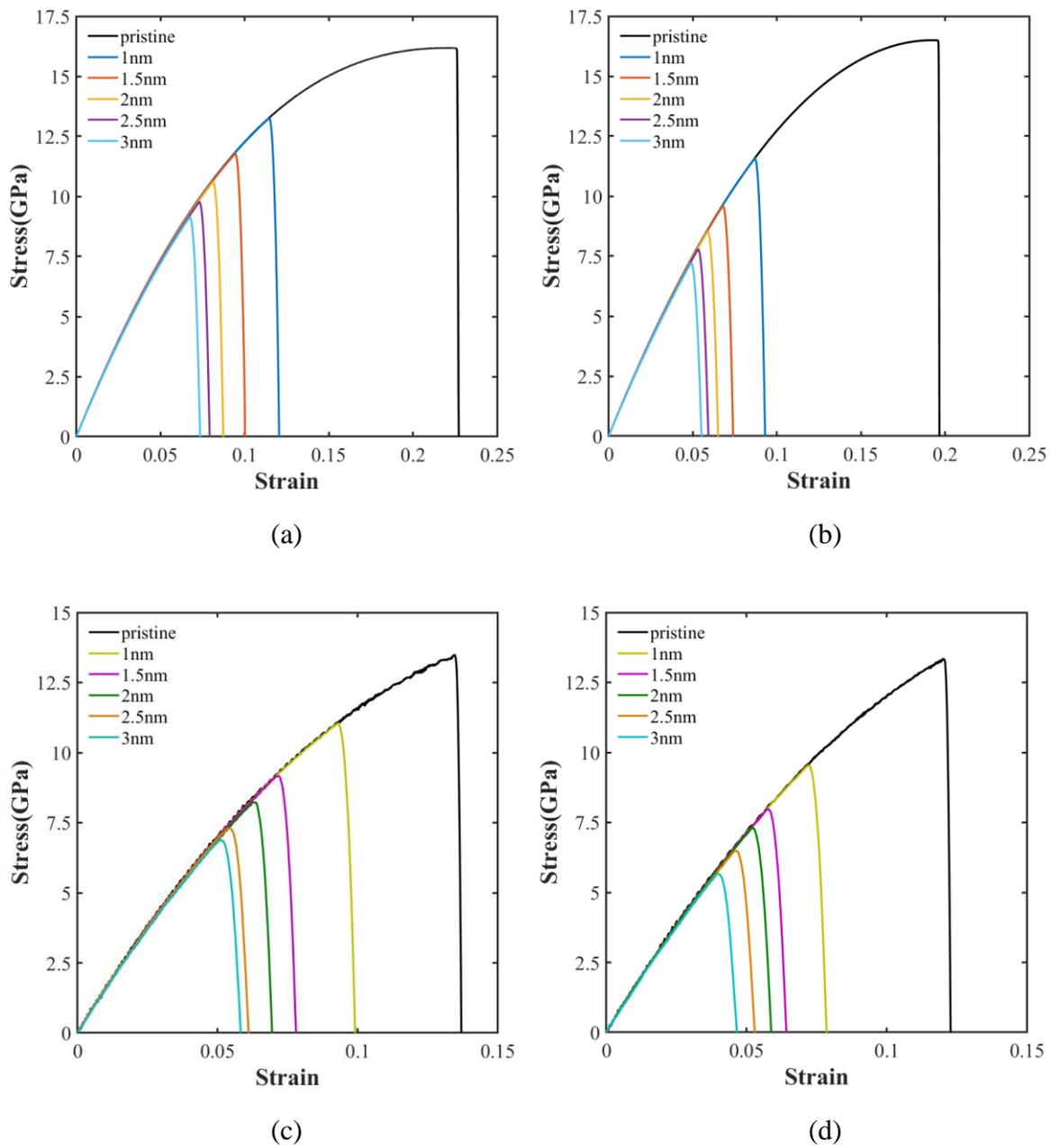

Figure S2- Stress-strain curve for nano-porous SLMoS$_2$ with the change of pore diameter (a) loading in zigzag direction at 1K (b) loading in armchair direction at 1K (c) loading in zigzag direction at 300K (d) loading in armchair direction at 300K

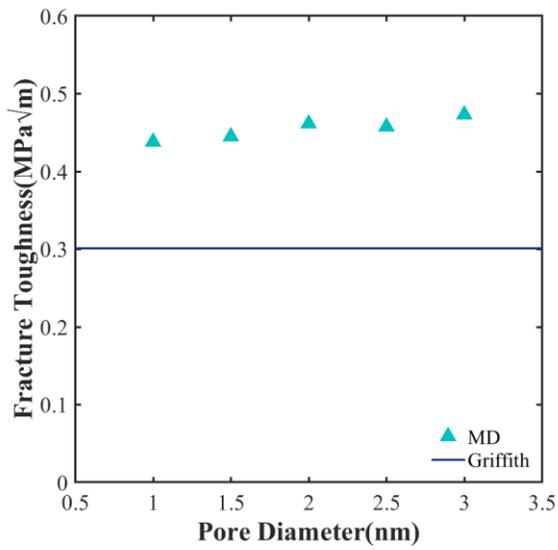
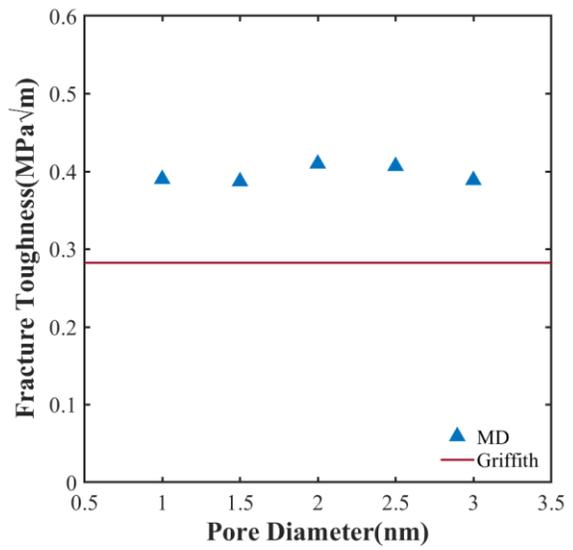

(a)            (b)

Figure S3- Variation of fracture toughness with the change of the initial pore diameter of SLMoS$_2$ under uniaxial tension along the (a) zigzag (b) armchair direction

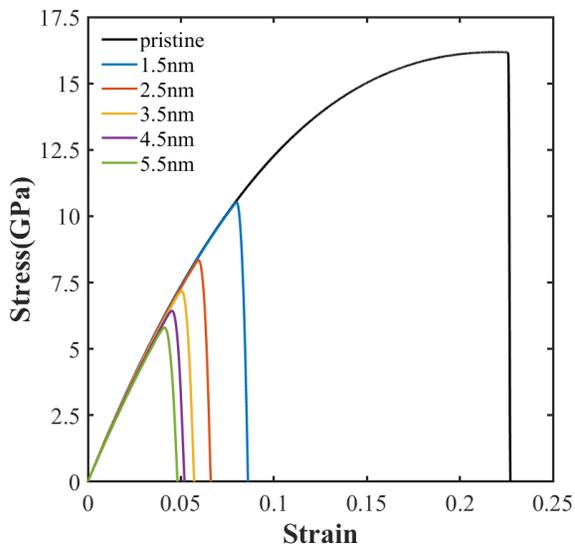
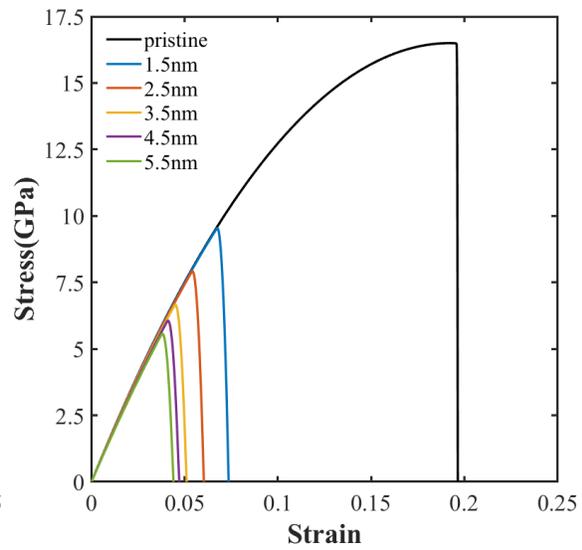

(a)            (b)

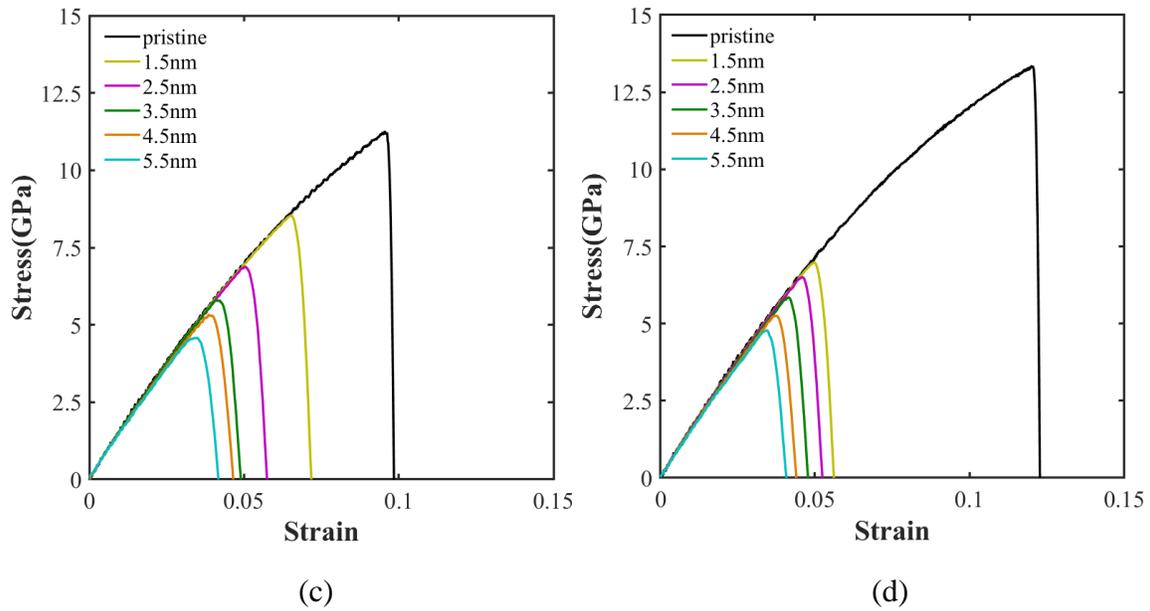

(c)   (d)

Figure S4- Stress-strain curve for crack length dependency for SLMoS$_2$ (a) crack in armchair direction at 1K (b) crack in zigzag direction at 1K (c) crack in armchair direction at 300K (d) crack in zigzag direction at 300K

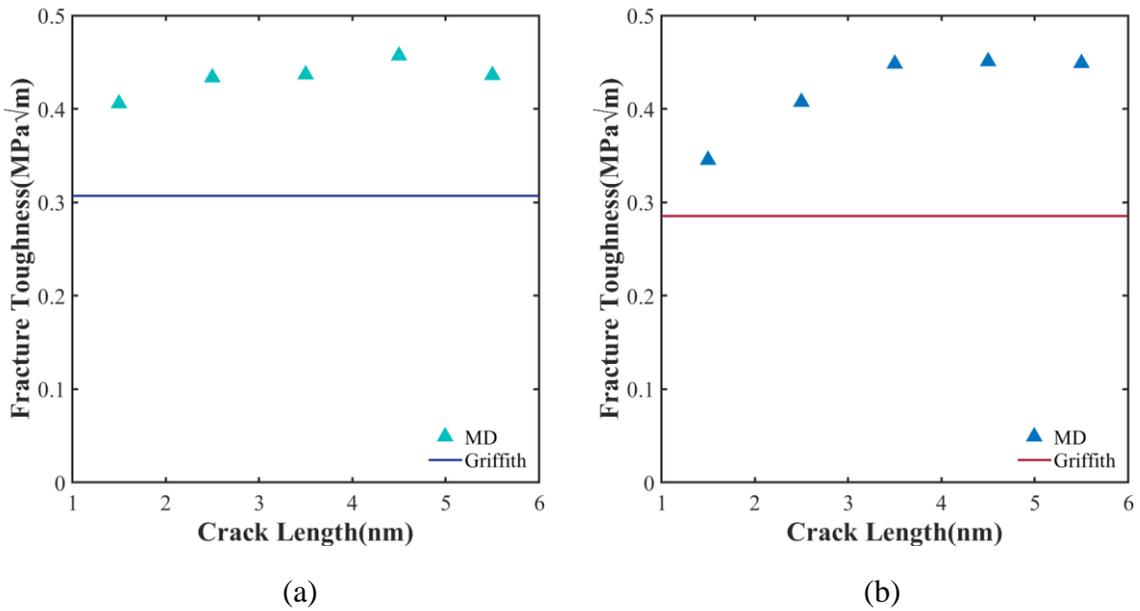

(a)   (b)

Figure S5- Variation of fracture toughness with the change of the initial crack length of SLMoS$_2$ under uniaxial tension along the (a) zigzag (b) armchair direction

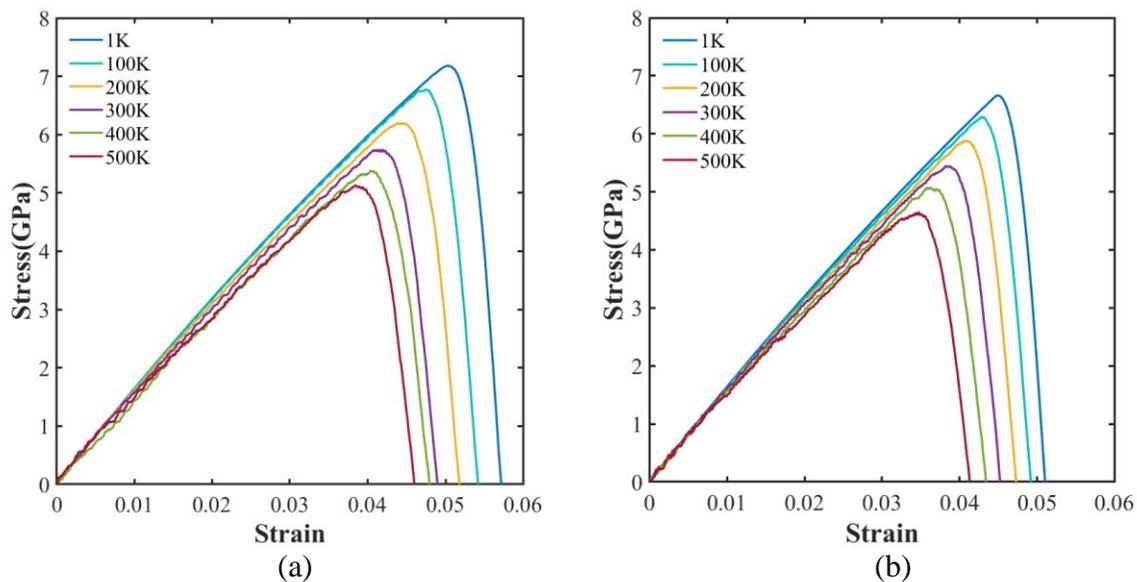

Figure S6- Temperature dependent stress-strain relationship of pre-cracked SLMoS2 having a fixed crack length of 3.5nm and strain rate $10^9$ s$^{-1}$ along (a) zigzag (b) armchair direction

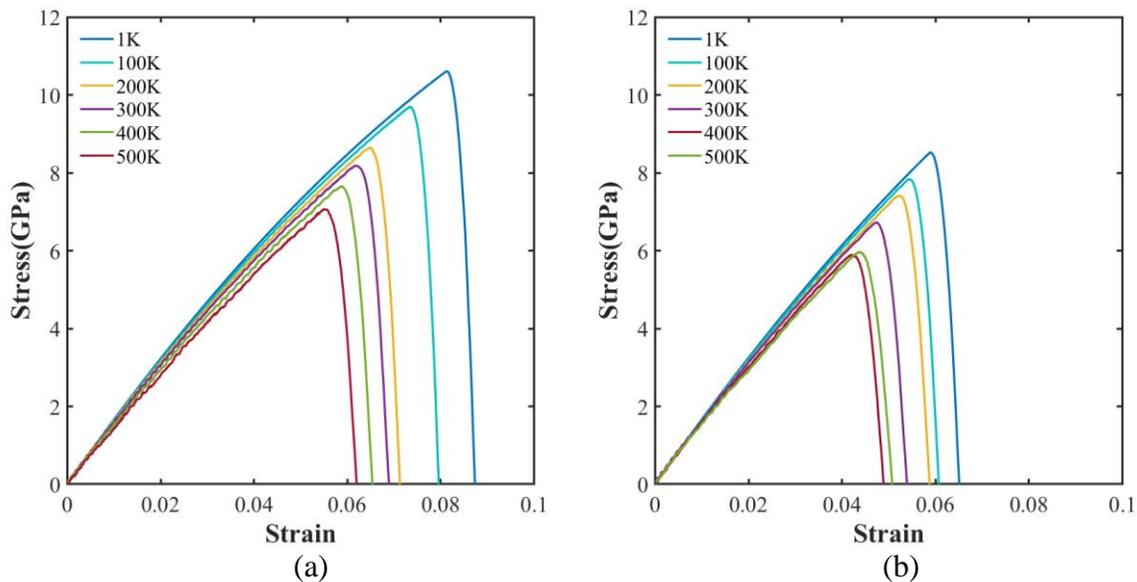

Figure S7- Temperature dependent stress-strain relationship of nano-porous SLMoS2 having a fixed pore dia. of 1nm and strain rate $10^9$ s$^{-1}$ along (a) zigzag (b) armchair direction

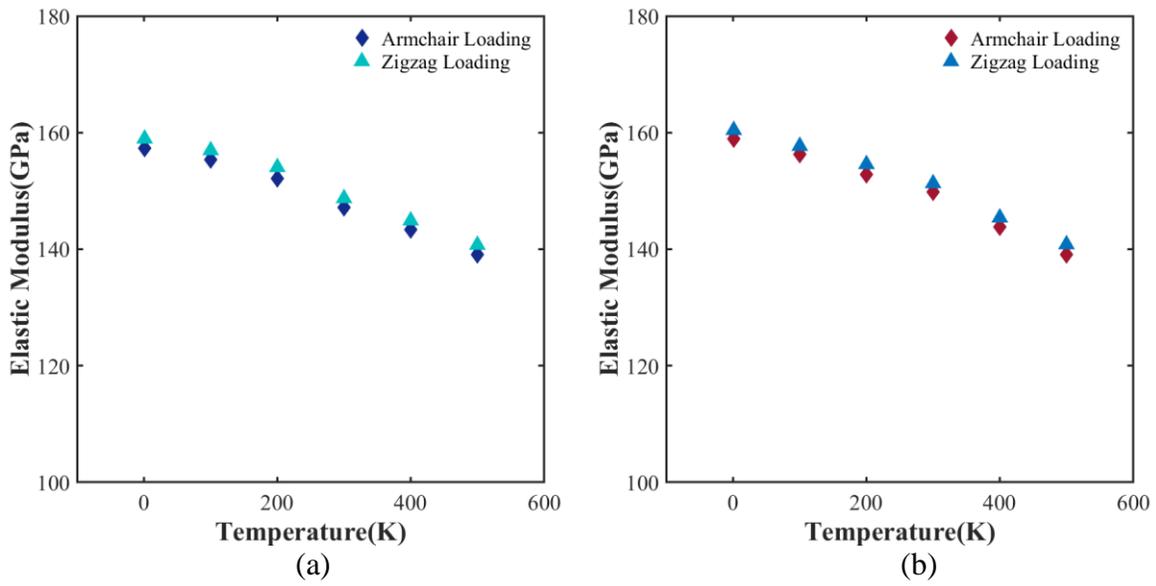

Figure S8- Variation of elastic modulus of (a) pre-cracked (b) nano-porous SLMoS$_2$ with the change of temperature

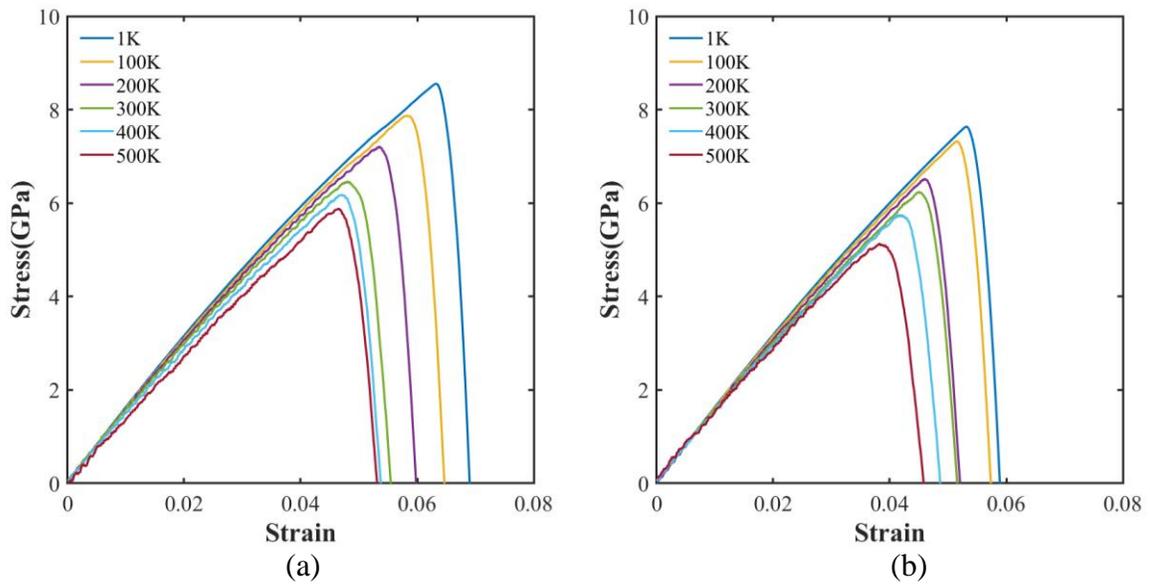

Figure S9- Temperature dependent stress-strain relationship of crack with pore SLMoS2 for a fixed strain rate $10^9$ s$^{-1}$ along (a) zigzag (b) armchair direction

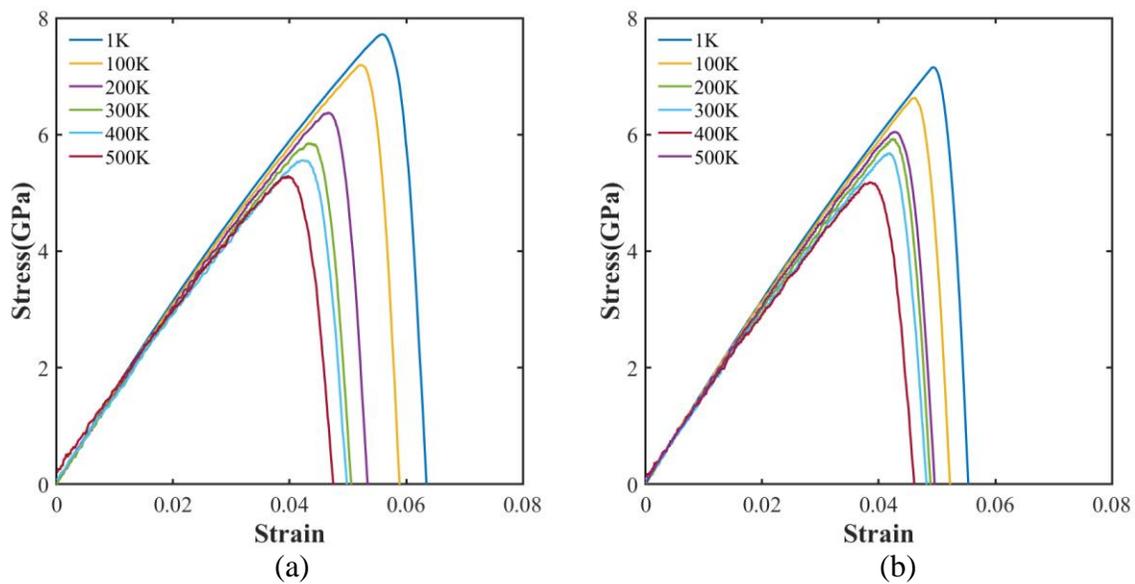

(a)                  (b)

Figure S10- Temperature dependent stress-strain relationship of main crack auxiliary crack combination of SLMoS2 for a constant strain rate $10^9$ s$^{-1}$ along (a) zigzag (b) armchair direction

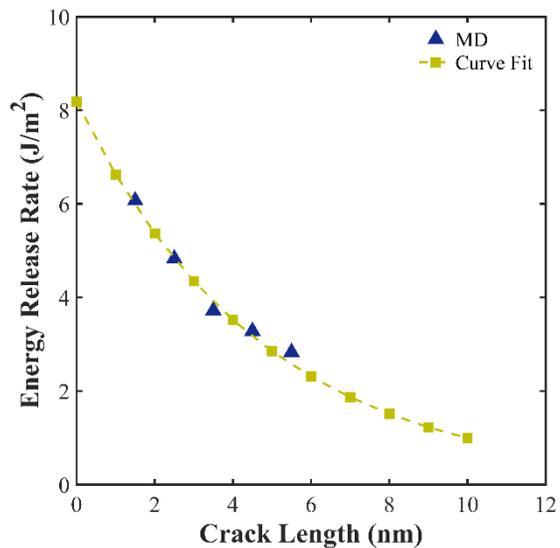

Figure S11- Energy release rate of MoS$_2$ monolayer for different crack lengths

Table S1- Pore and Auxiliary Crack Positions

| Position | Armchair (pore) | Zigzag (pore) | Armchair (Auxiliary crack) | Zigzag (Auxiliary Crack) | Position | Armchair (Auxiliary crack) | Position | Zigzag (Auxiliary Crack) |
|---|---|---|---|---|---|---|---|---|
| 1 | (155.39,160.61) | (10.80,9.36) | (158.45,176.20) | (174.66,159.06) | 1 | (158.45,176.21) | 2 | (174.66,162.18)) |
| 2 | (161.15,160.62) | (10.80,10.92) | (161.16,165.66) | (174.66,162.18) | 1(i) | (158.45,266.65) | 2(i) | (196.27,162.18) |
| 3 | (163.86,162.18) | (10.80,14.03) | (168.86,176.21) | (174.66,165.30) | 1(ii) | (158.45,219.87) | 2(ii) | (217.87,162.18) |
| 4 | (166.56,160.62) | (10.80,17.15) | (166.56,165.66) | (174.66,168.41) | 1(iii) | (158.45, 194.92) | 2(iii) | (239.48,162.18) |
| 5 | (169.26,159.06) | (10.80,20.27) | (169.26,176.21) | (174.66,171.51) | 1(iv) | (158.45, 179.33) | 2(iv) | (261.09,162.18 |
| 6 | (171.96,157.50) | (10.80,23.39) | | | | | | |

## SUPPLEMENTARY DISCUSSIONS

**Effect of temperature on Stress-Strain Relationship**

The effect of temperature on the stress-strain relationship of pre-cracked and nano-porous single layer $MoS_2$ has been studied accordingly for a fixed crack length of 3.5 nm and a pore diameter of 2nm pulled in both zigzag and armchair directions. Moreover, the temperature effect on the crack-pore and main crack auxiliary crack combination is shown in Figure 13 and Figure 14, respectively. Elevated temperature expedites the failure in all the cases. From Figures S6, S7, S9, S10, it is evident that for both zigzag and armchair loading conditions, the ultimate stress, elastic modulus, and failure strain decrease with the increase in temperature. This is because higher temperature facilitates higher fluctuations of bond lengths. This increases the probability of some bonds exceeding the critical bond length and initiates failure. Also, the increase in temperature causes higher entropy in the material which encourages crack propagation. As a result, weakening of the material occurs and the strength decreases. Again from Figure S8, it is clear that elastic modulus of the material always decreases as the temperature increases. And elastic modulus during ZZ loading remains greater than that of loading along AC direction. As bonds in armchair direction are much stretchier than that of zigzag directions. Therefore, bonds in armchair direction possess a better ability to resist crack propagation compared to zigzag direction.